\documentstyle[11pt,aaspp4]{article} 
\begin{document}
\title{A Progress Report on the Caltech Faint Galaxy Redshift Survey
\altaffilmark{1}}
\author{Judith G. Cohen\altaffilmark{2} }
\altaffiltext{1}{Based in large part on observations obtained at the
	W.M. Keck Observatory, which is operated jointly by the California 
	Institute of Technology and the University of California}
\altaffiltext{2}{Palomar Observatory, Mail Stop 105-24,
	California Institute of Technology, Pasadena, CA \, 91125}

\begin{abstract}
I review recent progress on determining the SEDs
and luminosity functions for galaxies in the 
large magnitude limited sample in the region of the HDF-North
of the Caltech Faint Galaxy Redshift Survey. 
\end{abstract}

\section{Spectral Energy Distributions and Galaxy Luminosity Functions}

The Caltech Faint Galaxy Redshift Survey, of which I am the PI,
has assembled a  redshift survey,
described in Cohen et al. (2000), of objects in the region of the HDF-N
observed with the LRIS (Oke et al. 1995) at the Keck Observatory 
over the past four 
years.   It includes 
redshifts for about 95\% of the sample of objects
in the photometric catalog of Hogg et al. (1999) with $R < 24$ within
the HDF itself, and for about 93\% of the sample of objects with 
$R < 23.5$ within
a circle of diameter 8 arcmin centered on the HDF, currently 
729 objects with measured $z$ in total. 
The analysis of this data set
has commenced, and several papers from our group have already been published.
Carlberg et al. (2000) discuss the kinematic pairs and their implications
for the merger rate, Hogg, Cohen
\& Blandford (2000) analyzes clustering of galaxies, and 
van den Bergh et al. (2000)
deals with the evolution of galaxy morphology, all to $z \sim 1$.  
We thus focus here 
on not yet published current and future work.

The logical next step is a derivation of the luminosity function for this
very complete faint sample of field galaxies.  We have utilized the 
photometric catalogs appropriate for
our HDF sample and used them to derive the rest frame SEDs of the galaxies.
In order to do this, one must adopt a model for the SED of a galaxy.  We adopt the
model of a double power law, whose index changes at 4000 \AA, for $F_{\nu}$.
While by no means a perfect representation of galaxy SEDs in their full
complexity, this simple model permits an illuminating comparison
of rest frame properties of galaxies.

A histogram of the two power law indices as a 
function of galaxy spectral type is
shown for our sample in the region of the HDF in figures 1a,b.  The galaxy 
spectral types are defined from our spectra
and refer to the relative dominance of emission or absorption lines in 
our galaxy spectra.  See  Cohen et al. (1999a) for details.

As expected, galaxies with signs of recent star formation
(i.e. those which show emission lines) have bluer continuum slopes in
the rest frame UV and the optical/near-infrared.

We see no change with redshift in the relationship between SED characteristics
and galaxy spectral type based on the strength of narrow emission and 
absorption features, except that actively star forming galaxies
become bluer in the mean in the rest-frame UV at higher redshift.
This regularity in the behavior of galaxy SEDs
across our broad wavelength coverage and up to
$z \sim 1$ is one reason why photometric redshifts work reasonably
well at least out to $z \sim 1$ with high precision photometric data sets.

At least some galaxy evolutionary synthesis models 
predict a 
more rapid change in SED parameters with redshift than is observed 
in our sample.  However, its not clear that the comparison is fair
in that the galaxy evolutionary models follow the evolution of
a particular galaxy with time, while we look at samples which we
believe to have similar spectra in various redshift bins.

The redder galaxies tend to be more luminous.  Although galaxies 
with strong absorption lines and no emission features
are $\sim$15\% of the total sample in the region of the HDF
with $0.25 < z < 0.8$, they are 
$\sim$50\% of the 25 most luminous galaxies in the sample at rest-frame $R$.

In many respects, these luminosity--density--spectrum correlations extend
results for rich clusters of galaxies to lower density environments.

Furthermore there is no evidence in our sample of a
population of very red galaxies with strong emission lines (i.e. dusty
starbursts).
Such objects would be detected if they fell within the magnitude
limit of the sample and their redshifts were such that the emission 
lines lie within the optical window.  
A detailed description of the SED analysis is given in Cohen (2000).


Figure 2 shows
the inferred rest frame luminosity at $K$ of our sample in the
region of the HDF.

Using standard maximum likelihood techniques and the formalism
described above, we have derived
the parameters of the best fitting luminosity function. 
The faint end slope of the luminosity function ($\alpha$)
for galaxies in the region of the HDF for $z \le 0.8$  shows
the same dependence of $\alpha$ on 
galaxy type as is seen locally.  Galaxies with strong
emission lines have a LF with a steeper $\alpha$.
The reddest galaxies have a much flatter low
luminosity slope.  The derived range of $\alpha$ is consistent
with local and other surveys, e.g. Lin et al. (1996) for the LCRS,
and Lin et al. (1999) for CNOC2.

The derived rest frame $L^*(R)$ is shown in figure 3
as a function of galaxy spectral type
and redshift.  Its behavior is consistent with other surveys at
lower $z$ and with the better determined (i.e. lower $z$) region
for the CFRS (Lilly et al. 1995).  Except  in
the  rest frame UV, it is also consistent with the
predictions obtained for passive galaxy evolution.
We have used  this formalism to calculate the evolution
of the LF over the full range of rest wavelengths and redshifts
included in our sample, but there is no room to discuss the
results here.
We have done the most detailed comparisons with the evolutionary
models of Poggianti (1997).
These seem to work well in the rest frame optical/near-IR, but not
so well in the rest frame UV, where problems might be anticipated.

Passive
evolution at constant stellar mass appears to be
a good approximation to the actual
behavior of at least the most luminous galaxies in
this large sample of galaxies in the region of the HDF
out to $z \sim 1.5$.   This applies to wavelengths redder than rest
frame $U$.

The total comoving number density for $L^*$ galaxies is
difficult to obtain for $z > 1$ from an optically selected
sample as there are many correction
factors which become larger with increasing redshift.
Pushing all of these, which include, for example, the
number of EROs missing from our sample,
as hard as possible, I deduce a comoving number density
for galaxies with $L > L^*/2$ for $z >1$ which is
80\% of the value we see for $z \sim 0.6$. 
Thus our cosmological density statistics and LFs
as a function of $z$ do not support
the suggestion of large scale merging of luminous galaxies near $z \sim 1$.
Substantial numbers of minor perturbative mergers are not ruled out.

\section{Signs of Galaxy Evolution}

The $z \sim 1$ ``$\cal E$'' galaxies are more luminous than their low $z$ counterparts.
We see, as have many other groups,
a strong evolution with $z$ of the prevalence and strength of emission lines, implying
that the mean SFR increases strongly between $z = 0$ and $z = 1$, by about a factor
of perhaps 10.  
The most luminous galaxies show this trend very strongly.  

We see modest evolution  of $L^*$ with $z$ consistent with
passive evolution.  We do not see strong changes
in the cosmological comoving volume density of luminous galaxies.
There is no evidence for a large increase with
$z$ of the rate of major mergers. 

To reconcile the above with the view that galaxies form or assemble near
$z \sim 1$ is hard.  
Most galaxies, and particularly the most
luminous galaxies, appear to have formed considerably earlier than $z \sim 1$
both in the field as well as in populous galaxy clusters.

\section{Starbursts and Other Future Work}

Using the spectroscopic definition of Cohen et al. (1999a), there
are 7 starbursts in the HDF sample. 
At least 3 of the 7 starbursts are
definite merger candidates in that there is a second galaxy in the sample within
a projected distance of 30 $h^{-1}$ kpc with exactly the same redshift
to within the measurement uncertainty.  

Future papers will present the equivalent widths of various absorption and
emission features in all the spectra, an analysis of galaxy abundances and star
formation rates based on this material, and details for the starbursts.  

\section{Acknowledgment}

I thank ESO for generous financial support towards attending this meeting.

\clearpage

\begin{figure}
\epsscale{0.7}
\plotone{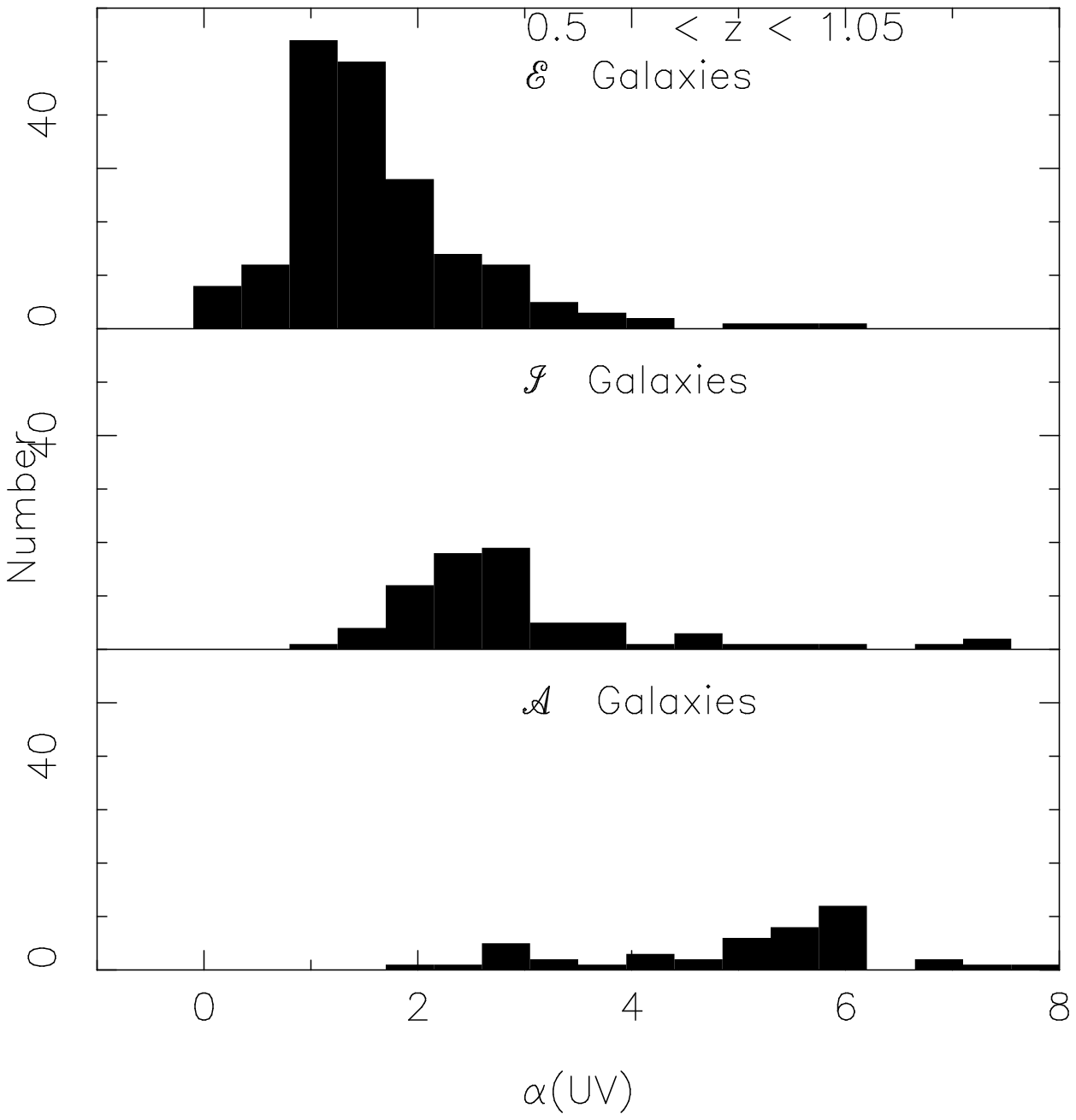}
\caption[]{A histogram of the the spectral index
$\alpha_{UV}$ is shown
for galaxies of spectral classes $\cal  A$, $\cal  I$ and $\cal  E$
in the range $0.5 < z < 1.05$.
\label{fig1a} }
\end{figure}

\clearpage

\begin{figure}
\epsscale{0.7}
\plotone{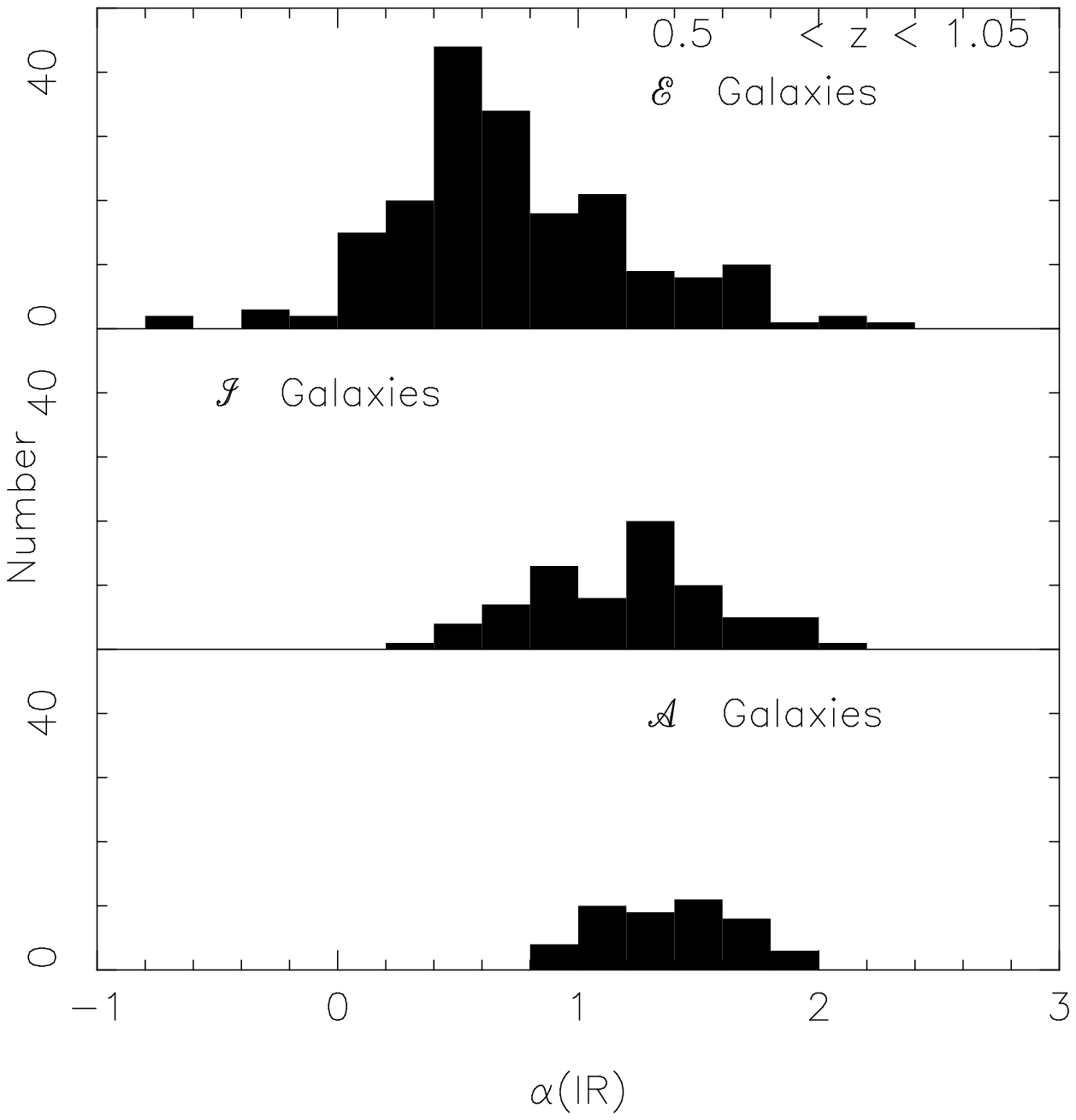}
\caption[]{A histogram of the the spectral index
$\alpha_{IR}$ is shown
for galaxies of spectral classes $\cal  A$, $\cal  I$ and $\cal  E$
in the range $0.5 < z < 1.05$.
\label{fig1b} }
\end{figure}

%
%
\clearpage

%
%
\begin{figure}
\epsscale{1.0}
\plotone{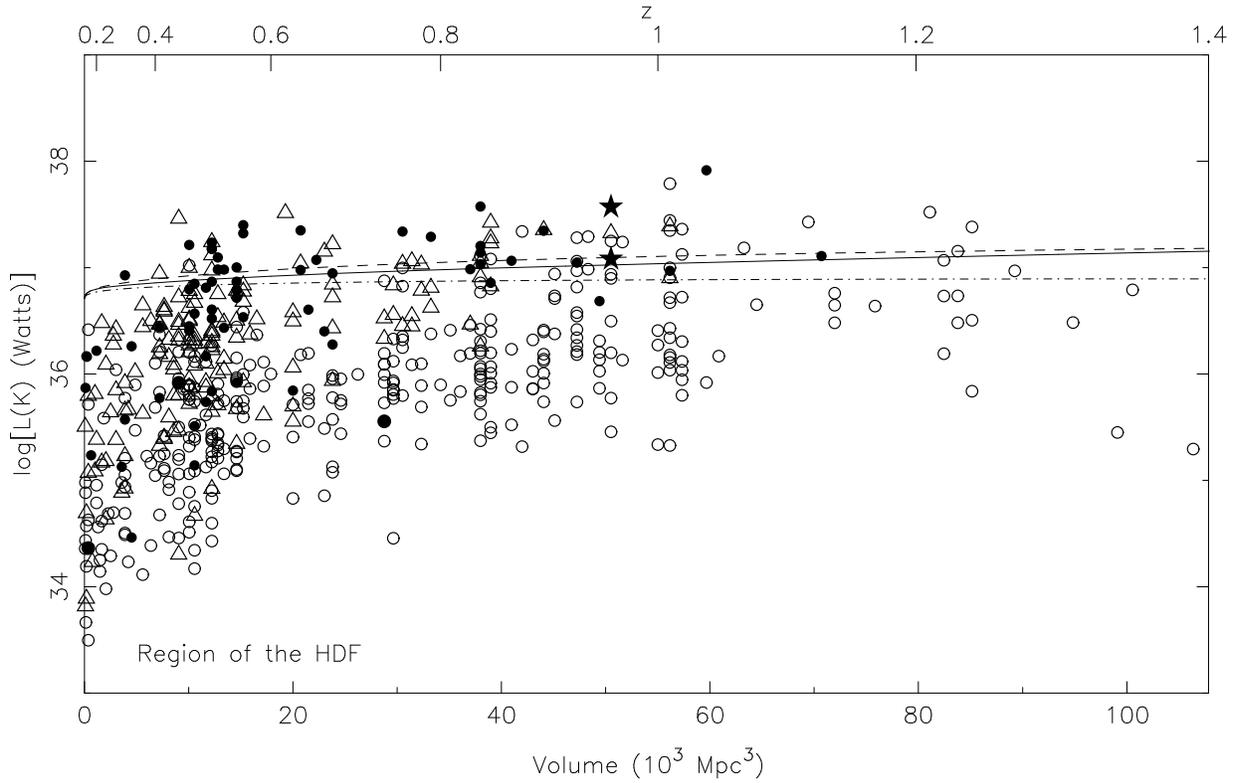}
\caption[]{$L(K)$ is shown as a function of the
cosmological comoving volume for galaxies
in the region of the HDF with secure redshifts.  The lines at the top
of the distribution represent the track of a galaxy of
a $L^*$ (at $K$) galaxy
with $L = 10^{11} L_{\odot}$ at $z=0$.  The
evolutionary corrections at $K$ calculated by
Poggianti (1997)
for E, Sa and Sc galaxies are applied in calculating these tracks for $z>0$.}
\label{fig2}
\end{figure}

\begin{figure}
\epsscale{0.9}
\plotone{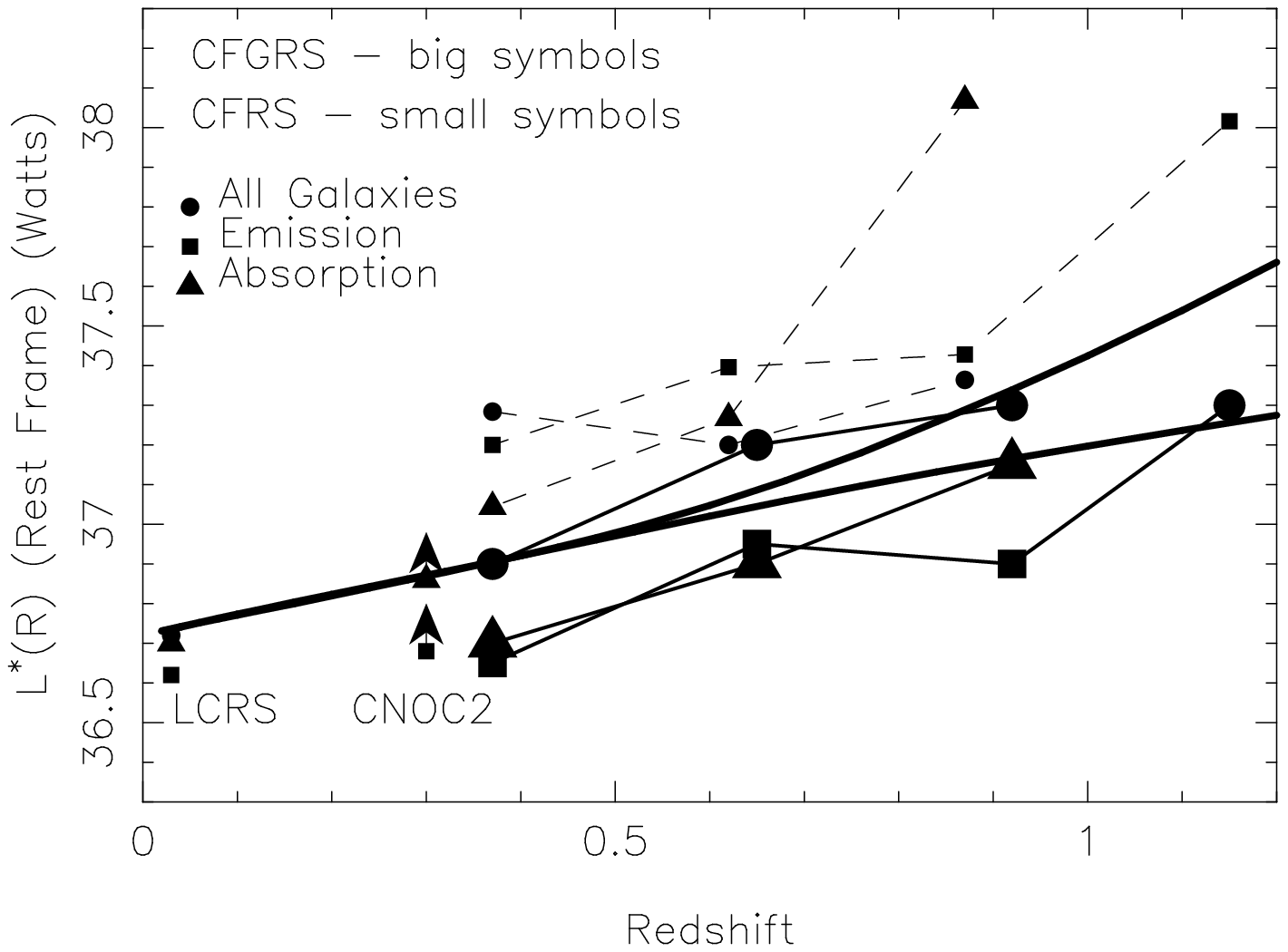}
\caption[]{The rest frame luminosity at $R$ is shown
as a function of redshift for data from the LCRS, CNOC2,
CFRS and the present sample.   The thick solid lines represent the 
track of a passively evolving elliptical or Sc galaxy calculated by
Poggianti (1997) set to the local
value of $L^*(R)$ at $z=0$.}
\label{fig3}
\end{figure}



\begin{thebibliography}{}
%
\addcontentsline{toc}{section}{References}


\bibitem{} Carlberg, R.~G., Cohen, J.~G., Patton, D.~R., et al.,
2000, ApJ, 532, L1

\bibitem{} Cohen, J.~G.,  Blandford, R., Hogg, D.~W., Pahre, M.~A. \&
Shopbell, P.~L., 1999a, ApJ, 512, 30

\bibitem{} Cohen, J.~G., Hogg, D.~W., Pahre, M.~A., Blandford, R., 
Shopbell, P.~L. \& Richberg, K., 1999b, ApJS, 120, 171


\bibitem{} Cohen, J.~G., Hogg, D.~W., Blandford, R., Cowie, L.~L.,
Hu, E., Songaila, A., Shopbell, P. \& Richberg, K., 2000, ApJ, 538, 29

\bibitem{} Cohen, J.~G., 2000, submitted to the AJ

\bibitem{} Cohen, J.~G., 2001, manuscript in preparation

\bibitem{} Hogg, D. W., Cohen J. G. \& Blandford R., 2000, ApJ (in press)

\bibitem{} Hogg, D.~W., Pahre M.~A., Adelberger K.~L., Blandford R., Cohen J.~G.,
Gautier T.~N., Jarrett T., Neugebauer G. \& Steidel C.~C., 2000, 
ApJS, 127, 1 (H00)

\bibitem{} Lilly, S.J., Tresse, L., Hammer, F.,  Crampton, D. \& LeFe\'vre, O.,
1995, ApJ, 455, 108

\bibitem{} Lin, H., Kirshner, R.P., Shectman, S.A., Landy, S.D., Oemler, A.,
Tucker, D.L. \& Shechter, P.L., 1996, ApJ, 464, 60

\bibitem{} Lin. H., Yee, H.K.C., Carlberg, R.G., et al., 1999, ApJ, 518, 533


\bibitem{} Oke, J.~B.,  Cohen, J.~G., Carr, M., et al., 1995, PASP,
107, 307

\bibitem{} Poggianti, B.~M., 1997, A\&A Supl, 122, 399

\bibitem{} van den Bergh, S., Cohen, J.~G., Hogg, , D.~W. \& Blandford, R.,
2000, AJ, 120, 2190
\end{thebibliography}
\end{document}